\documentclass[12pt]{iopart}

\usepackage{graphicx} 
\usepackage{bbm}
\usepackage{amssymb}

\begin{document}

\title[Grain boundary diffusion in a Peierls-Nabarro potential]{Grain boundary diffusion in a Peierls-Nabarro potential}

\author{F Leoni$^1$ and S Zapperi$^{2,3}$}
\address{$^1$ Dipartimento di Fisica, Sapienza --- Universit\`a di Roma, P.le
  A. Moro 2, 00185 Roma, Italy}
\address{$^2$ CNR-INFM, SMC, Dipartimento di Fisica, Sapienza --- Universit\`a di Roma, P.le
  A. Moro 2, 00185 Roma, Italy}
\address{$^3$ ISI Foundation, Viale S. Severo 65, 10133 Torino, Italy }
\ead{leoni@pil.phys.uniroma1.it}

\begin{abstract}
We investigate the diffusion of a grain boundary in a crystalline material. 
We consider in particular the case of a regularly spaced low-angle grain boundary
schematized as an array of dislocations that interact with each other through 
long-range stress fields and with the crystalline Peierls-Nabarro potential. 
The methodology employed to analyze the dynamics of the center of mass of the 
grain boundary and its spatio-temporal fluctuations is based on over-damped Langevin 
equations. The generality and the efficiency of this technique is proved by the 
agreement with molecular dynamics simulations.
\end{abstract}

\maketitle

\section{Introduction}

Understanding interface kinetics in materials is an important theoretical and
practical problem, since this process influences the microstructure, such as
the grain size, the texture, and the interface type. 
From the theoretical point of view, the study of processes that involve
surface and interface properties gained significant interest in
non-equilibrium statistical mechanics \cite{barabasi,krug}. In particular,
dislocations \cite{zaiser,zapperi} and grain boundaries \cite{moretti,moretti2}
provide a concrete example of driven elastic manifolds in random
media \cite{kardar}. Other example of this general problem are domain walls in
ferromagnets \cite{lemerle,zapperi2}, flux line in type II superconductors
\cite{bhattacharya,surdeanu}, contact lines \cite{schaffer,rolley} and crack fronts
\cite{bouchaud,schmittbuhl}.  From the point of view of applications,
understanding grain boundary kinetics has a great importance for
polycrystalline materials, since the resulting grain microstructure 
determines material properties such as strength,
hardness, resistance to corrosion, conductivity etc. \cite{sutton}. 
Hence the ambitious goal of these studies is to be able to control 
the microstructural properties of polycrystals. 

Several approaches have been employed in the literature to study
grain boundary kinetics. Ref.~\cite{trautt} employs molecular dynamics (MD)
simulations with appropriate interatomic interactions to study the diffusion
of grain boundaries at the atomic scale \cite{trautt}. The method allows
to quantify the mobility of grain boundaries and to compare the
results with experiments \cite{trautt}. While MD simulations provide
a very accurate description of the dynamics, the method suffers from numerical limitations 
and it is difficult to reach the asymptotic regime. An alternative method is provided
by the Langevin approach in which the grain boundary
is assumed to evolve stochastically in an external  potential \cite{risken}.
The dynamics of the underlying crystalline medium enters in the problem only through the
noise term (due to lattice vibrations) and the periodic potential (Peierls-Nabarro).
Hence, the equations of motion of the atoms or molecules are not directly relevant.
Indeed there is experimental evidence in supporting of separation of time scales
in plastic flow \cite{ananthakrishna} and it is thus possible to integrate out the fast degrees of
freedom (atomic vibrations) and consider only the slow ones (dislocations position). 

Here we study the evolution of a grain boundary (GB) in a crystalline
material by the Langevin approach. The GB is treated as an array of interacting
dislocations performing a thermally activated motion in a periodic (Peierls-Nabarro)
potential. Similar models have been employed in the past to 
study the conductivity of superionic conductors \cite{fulde,dieterich}, the relaxational dynamics of
rotators \cite{marchesoni} and Josephson tunneling junctions \cite{risken}. Notice
that the crucial role  played by long-range stresses is often disregarded
in analyzing GB deformation. On the other hand, it has been shown in Ref.~\cite{moretti} that
a surface tension approximation for the GB stiffness is inappropriate and
one has to consider explicitly non-local interactions. The present model incorporates
this effect in the equations of motion.

We simulate the set of Langevin equations numerically to describe
the GB kinetics and its fluctuations. The results are in good agreement
with MD simulations \cite{trautt} and allow to clarify the origin of the
short time deviations from the diffusive behavior observed in Ref.~\cite{trautt}. 
In addition, a linearized version of the model can be treated analytically and the asymptotic results
are found in good agreement with the simulations.
The manuscript is organized as follows: in Sec. II we introduce the model, which is first studied
in the flat GB limit in Sec. III. Sec. IV presents numerical simulations of the full 
flexible GB problems and Sec. V discusses the continuum theory. Sec. VI is devoted
to conclusions.

\section{The model}

To study the GB dynamics we consider a phenomenological mesoscopic
approach. We consider in particular the case of a regularly spaced low-angle 
grain boundary schematized as an array of straight dislocations that interact with each other
through long-range stress fields and with the crystalline Peierls-Nabarro (PN) potential. 
The GB is composed by $N$ dislocations where configurations are repeated {\it
ad infinitum} because of periodic boundary conditions along the $y$ direction. 
Each dislocation has Burger vector of modulus $b$ parallel to the $x$ axis and
the distance between two adjacent dislocations along the $y$ direction is 
fixed to be $a$. 
Each straight dislocation interacts with the lattice and with others
dislocations through long-range stress fields. 
The effect of the lattice over each n-th dislocation can be decomposed as the
sum of three contributions:\\
\begin{itemize}
\item $F_{PN}(x_n)=-\mathcal{A}\frac{\mbox{\normalsize{$\mu
  b$}}}{\mbox{\normalsize{$2\pi
  r_0$}}}\sin(\frac{\mbox{\normalsize{$2\pi
  x_n$}}}{\mbox{\normalsize{$b$}}})$, the PN force where $\mathcal{A}$
  is the area of the GB, $\mu$ is the shear modulus and $r_0$ the
  inter-atomic distance;
\item -$\gamma \dot{x}_n(t)$, the average effect of the lattice fluctuations
  where $\gamma$ is the viscosity coefficient;  
\item $\gamma\eta_n(t)$, the impulsive effect of the lattice fluctuations
  assumed to be Gaussian for the central limit theorem and uncorrelated in
  space and time: $\langle\eta_n(t)\rangle=0$,
  $\langle\eta_n(t)\eta_m(t')\rangle=D\delta_{nm}\delta(t-t')$ where $D$ is
  the diffusion coefficient \cite{risken}.
\end{itemize}
\vspace{0.5cm}
The long-range stress field exercised by all the other dislocations over the
n-th, the Peach-Koehler force $F_{PK}^{n,N}({\bf x},{\bf y})$, is computed
considering the image dislocations method to comply with periodic boundary
conditions along the $y$ direction. Making use of calculations in
\cite{hirth,friedel} one can find the following expression
\begin{equation}
\begin{array}{l}
\fl F_{PK}^{n,N}({\bf x},{\bf y})= -\frac{\mbox{\normalsize{$\mu
  b^2\pi$}}}{\mbox{\normalsize{$N^2a^2(1-\nu)$}}}\sum_{m=1}^{N}(x_n-x_m)\cdot\\
\\
\cdot\frac{\mbox{\normalsize{$\{\cosh[2\pi(x_n-x_m)/Na]\cos[2\pi(y_n-y_m)/Na]-1\}$}}}
  {\mbox{\normalsize{$\{\cosh[2\pi(x_n-x_m)/Na]-\cos[2\pi(y_n-y_m)/Na]\}^2$}}},
\end{array}
\end{equation}
\vspace{0.5cm}
where $\nu$ is the Poisson's ratio, $y_n=n\cdot a$ and $y_m=m\cdot a$.
Finally the over-damped Langevin equation \cite{risken} for the GB reads
\begin{equation}\label{general_GB}
\gamma\dot{x}_n(t)=F_{PN}(x_n)+F_{PK}^{n,N}({\bf x},{\bf y})+\gamma\eta_n(t) ,
\end{equation}
for $n=1,...,N$, or rather
\begin{equation}\label{flexible_GB}
\begin{array}{l}
\fl \dot{x}_n(t) =  -\mathcal{A}\frac{\mbox{\normalsize{$\mu
  b$}}}{\mbox{\normalsize{$2\pi r_0\gamma$}}}\sin(\frac{\mbox{\normalsize{$2\pi
  x_n$}}}{\mbox{\normalsize{$b$}}})-\frac{\mbox{\normalsize{$\mu
  b^2\pi$}}}{\mbox{\normalsize{$N^2a^2(1-\nu)\gamma$}}}\sum_{m=1}^{N}(x_n-x_m)\cdot\\
\\
\cdot\frac{\mbox{\normalsize{$\{\cosh[2\pi(x_n-x_m)/Na]\cos[2\pi(y_n-y_m)/Na]-1\}$}}}
  {\mbox{\normalsize{$\{\cosh[2\pi(x_n-x_m)/Na]-\cos[2\pi(y_n-y_m)/Na]\}^2$}}}+\eta_n(t) .
\end{array}
\end{equation}
\vspace{0.5cm}
To indicate the amplitude of the $F_{PN}$ and $F_{PK}$ forces we introduce
respectively the parameters $A_{PN}=\mathcal{A}\mu b/2\pi r_0\gamma$ and
$A_{PK}=\mu b^2\pi/a^2(1-\nu)\gamma$. 

The key quantities that we consider in order to characterize the dynamics of the GB are:
\begin{itemize}
\item the mean-square displacement of the center of mass, $\Delta
  x_{cm}(t)=\langle x_{cm}^2(t)\rangle-\langle x_{cm}(t)\rangle^2=\langle
  \overline{x_n(t)}^2\rangle-\langle \overline{x_n(t)}\rangle^2$ where 
  $x_{cm}(t)=\overline{x_n}=1/N\sum_{n=1}^{N}x_n(t)$;
\item the mean-square width $W^2(t)=\overline{\langle
  x_n^2\rangle}-\langle(\overline{x_n})^2\rangle$.
\end{itemize}

In the following, we first analyze the case of a flat GB for which a comparison with
MD simulations approach \cite{trautt} is made. Next we consider the full flexible description of the
GB. Finally, we discuss a linearized version of the model that can be treated analytically.

\section{Flat grain boundary} 

For many applications a good approximation is to consider a flat GB with a
single degree of freedom, for which $F_{PK}^{n,L}({\bf x},{\bf y})=0$ and
$x_n(t)=x_{cm}(t)$ for $n=1,...,N$. In other words, the flat GB is described by
the following equation
\begin{equation}\label{flat_GB}
\dot{x}_{cm}(t)=-\mathcal{A}\frac{\mu b}{2\pi r_0\gamma}\sin(\frac{2\pi
  x_{cm}}{b})+\eta(t) ,
\end{equation}
where the correlation properties of the thermal fluctuations are:
$\langle\eta(t)\rangle=0$ and $\langle\eta(t)\eta(t')\rangle=D\delta(t-t')$.
This type of equation, also known as the Kramers equation with periodic
potential, has been extensively studied in the literature \cite{risken}. In particular, 
the mean-square displacement is known to display a combination of 
oscillatory and diffusive behavior
\cite{risken,ferrando}. Different dynamical regimes are found as the potential
strength or the friction varies \cite{ferrando}.
In fact, we show next that this simple model allows to understand the 
short-time deviations from diffusive behavior observed in MD
simulations \cite{trautt}.

\begin{figure}[h!]
\begin{center}
\includegraphics[clip=true,width=10cm]{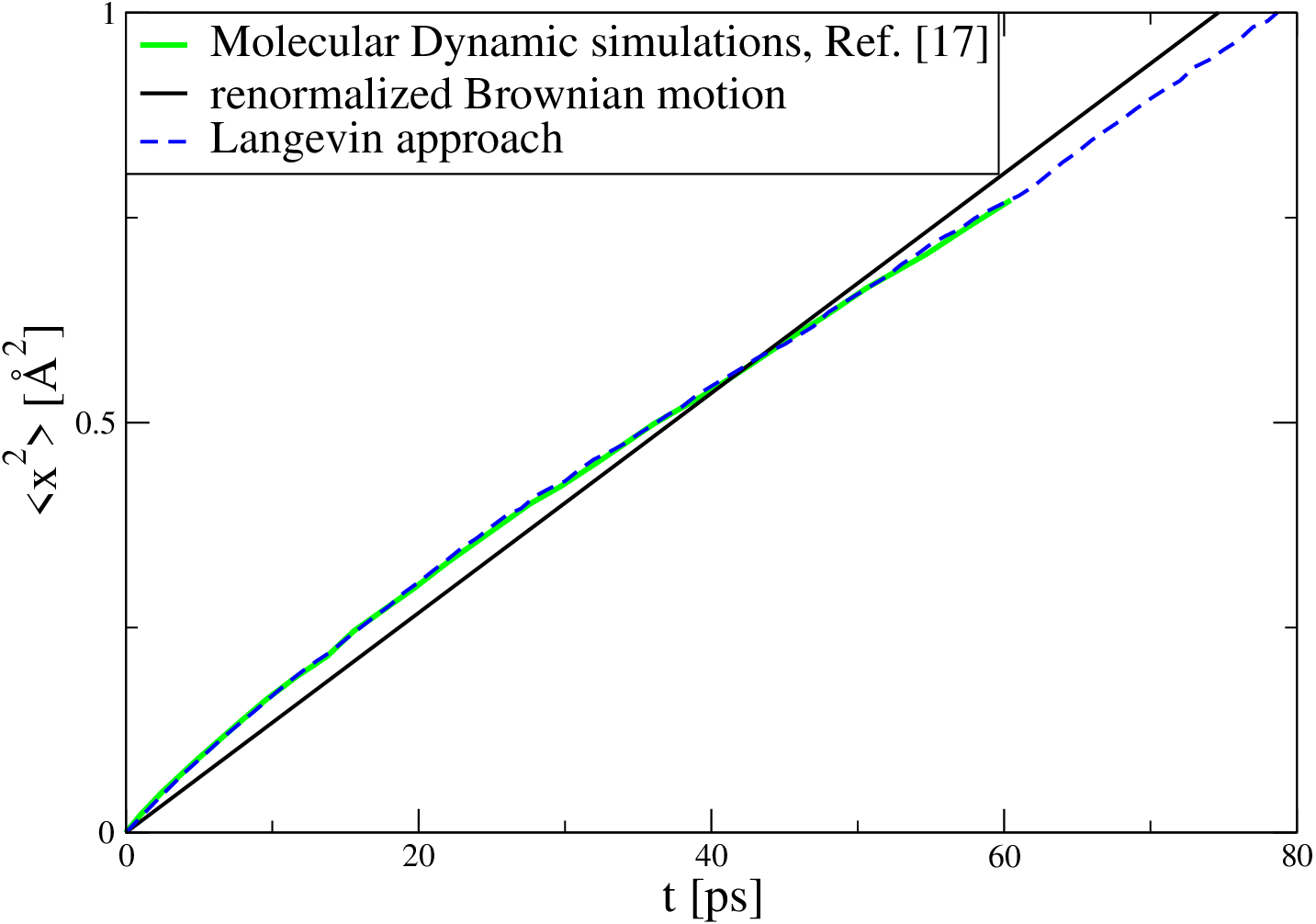}
\end{center}
\caption{Mean-square displacement $\Delta x_{cm}(t)$ of the flat grain boundary 
comparison between molecular dynamic simulation \cite{trautt} (green line)
and Langevin approach simulation (dashed blue line). Both type of simulations
predicts a linear dependence in time of $\Delta x_{cm}(t)$ for long times
represented in the figure by the renormalized free Brownian motion (straight
black line).}
\label{graphic1} 
\end{figure}

Integrating Eq.~\ref{flat_GB} with the initial condition $x_{cm}(0)=0$ by
means of computer simulations (fitting $D$ and $\gamma$ with the condition $D\gamma=D_R/\mathcal{M}$
\cite{risken} where $\mathcal{M}$ is the mobility and $D_R$ is the
renormalized diffusion coefficient considered in \cite{trautt}), we
have compared the mean-square displacement $\Delta x_{cm}(t)$ to the
one obtained from MD simulation in Ref.~\cite{trautt}.
In Fig.~\ref{graphic1} this comparison is displayed together with the
mean-square displacement of the renormalized free Brownian motion (described by the
equation: $\dot{x}=\eta(t)$ with $\langle\eta(t)\rangle=0$ and
$\langle\eta(t)\eta(t')\rangle=D_R\delta(t-t')$).  
The agreement between the two simulations is extremely good. For higher
times ($t>80ps$), the mean-square displacement tends to the renormalized
Brownian motion. Hence taking explicitly into account the sinusoidal Peierls-Nabarro
force in the Langevin equation allows to describe the mean-square displacement
for early times of the dynamics. 

One can deduce in a simplified intuitive way the temporal evolution of the
mean-square displacement starting by the transition probability density $P$
for small times (small $\tau$) \cite{risken}  
\begin{equation}
P(x,t+\tau|x',t)=\frac{1}{2\sqrt{\pi D\tau}}\ \
\mbox{\Large{$e$}}^{\mbox{\large{$-\frac{[x-x'-F_{PN}(x)\tau]^2}{4D\tau}$}}}.
\end{equation}
Next we consider the transition probability $P_{01}(x_0+\Delta
x_1,t_0+\tau|x_0,t_0)$ to run from the point $x_0$ at time $t_0$ to the point
$x_1=x_0+\Delta x_1$ at time $t_1=t_0+\tau$ and $P_{12}(x_0+\Delta x_1+\Delta
x_2,t_0+2\tau|x_0+\Delta x_1,t_0+\tau)$ to run from $x_1$ at $t_1$ to
$x_2=x_1+\Delta x_2$ at $t_2=t_1+\tau$
\begin{equation}\left\{\begin{array}{rll}
P_{01} & = & \mbox{\Large{$\frac{1}{2\sqrt{\pi D\tau}}$}} \ \
\mbox{\Large{$e$}}^{\mbox{\large{$-\frac{[\Delta x_1-F_{PN}(x_0)\tau]^2}{4D\tau}$}}}\\
&&\\
P_{12} & = & \mbox{\Large{$\frac{1}{2\sqrt{\pi D\tau}}$}} \ \
\mbox{\Large{$e$}}^{\mbox{\large{$-\frac{[\Delta x_2-F_{PN}(x_0+\Delta x_1)\tau]^2}{4D\tau}$}}}.\\
\end{array}\right.
\end{equation}
For a free Brownian motion ($F_{PN}(x)=0$) the condition $P_{01}=P_{12}$ implies
$\Delta x_1=\Delta x_2$ and stochastic displacements are space
independent. If we impose this condition in presence of a periodic
force  $F_{PN}(x)=-dU_{PN}(x)/dx$, we obtain
\begin{equation}\begin{array}{l}
P_{01}=P_{12} \ \ \Rightarrow \ \ \Delta x_1-F_{PN}(x_0)\tau=\Delta
x_2-F_{PN}(x_0+\Delta x_1)\tau \ \ \Rightarrow\\
\\
\Rightarrow \ \ \Delta x_2=\Delta
x_1\left[1+\left.\mbox{\large{$\frac{dF_{PN}(x)}{dx}$}}\right|_{x_0}\tau\right] ,
\end{array}
\end{equation}
and then
\begin{equation} 
\Delta x_2 \gtrless\Delta x_1 \ \ \ \ \mbox{if} \ \ \ \
\left.\frac{dF_{PN}}{dx}\right|_{x_0}\gtrless 0 \ \ \ \ \mbox{or rather} \ \ \
\ \left.\frac{d^2U_{PN}}{dx^2}\right|_{x_0}\lessgtr 0 .
\end{equation}
This result implies that, with the initial condition $x_{cm}(0)=0$, if the
potential $U_{PN}(x)$ is convex (concave) the mean-square
displacement curve is concave (convex). In the case of the PN
potential, we find indeed that the mean-square displacement curve should display upper and
lower deviations from the straight line, corresponding to a renormalized free Brownian motion,
depending in $d^2U_{PN}(x)/dx^2$. These deviations decrease with time so that for
large times the curve should approach a straight line \cite{risken}. 

\section{Flexible grain boundary} 

A more general description of the GB considers its internal deformation and the
dynamics is described by Eq.~\ref{flexible_GB}. The dynamical
behavior of the GB depends on the amplitude of the three terms in the
right-hand side of Eq.~\ref{flexible_GB}. 
\begin{figure}[h!]
\begin{center}
\includegraphics[clip=true,width=10cm]{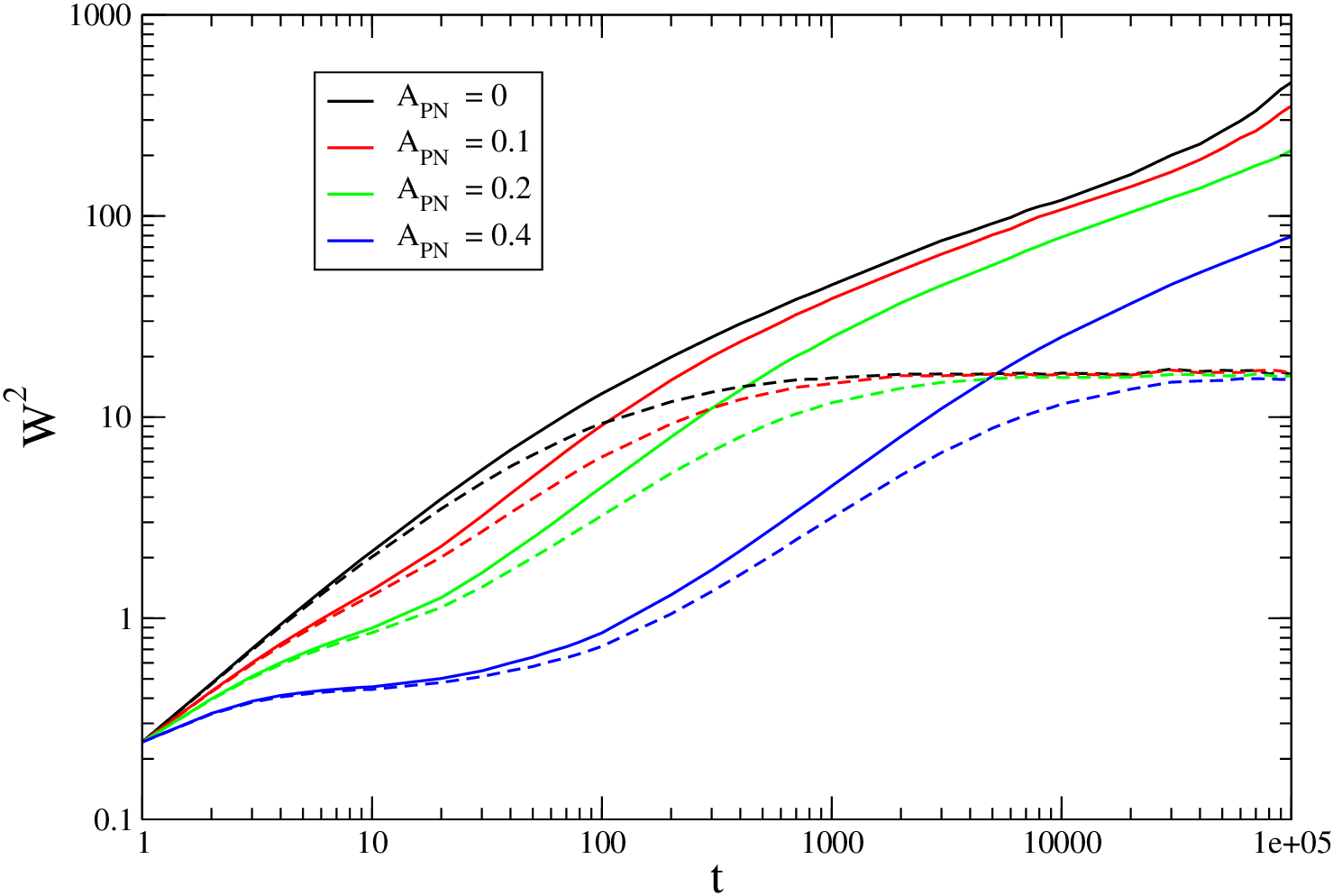}
\end{center}
\caption{Mean-square width of the grain boundary, $W^2(t)$, in the case of
  $N=32$ for the two typical situation in which (continuum line) the GB
  exfoliate ($W^2(t)$ increase with time) because the noise is high enough
  respect to $A_{PK}$ and to $A_{PN}$, (dashed line) the GB reach a
  stationary state ($W^2(t)$ saturates after a certain time) because the noise is
  small enough respect to $A_{PK}$ or to $A_{PN}$.}
\label{graphic8} 
\end{figure}
The parameters that characterize the behavior of the GB are $a$, $b$,
$A_{PK}$, $A_{PN}$ and $D$. Varying the values of these parameters, in the
long-time limit the GB can either exfoliate (when the noise, $D$, is high enough
with respect to $A_{PK}$ and to $A_{PN}$) or reach a stationary state (when
the noise, $D$, is small when compared to $A_{PK}$ or to $A_{PN}$). The asymptotic
behavior can be read off from the width $W^2(t)$ that keeps on increasing when
the GB exfoliates and saturates when the GB remains stable. In
Fig.~\ref{graphic8} the comparison between these two typical situations is
displayed in the case of $N=32$, $a=b=3\pi$, $A_{PN}=0,0.1,0.2,0.4$, $D=0.25$
and $A_{PK}=0.14$ for the case in which the GB remains stable, while
$A_{PK}=0.0896$ for the case in which the GB exfoliates. 

\begin{figure}[h!]
\begin{center}
\includegraphics[clip=true,width=10cm]{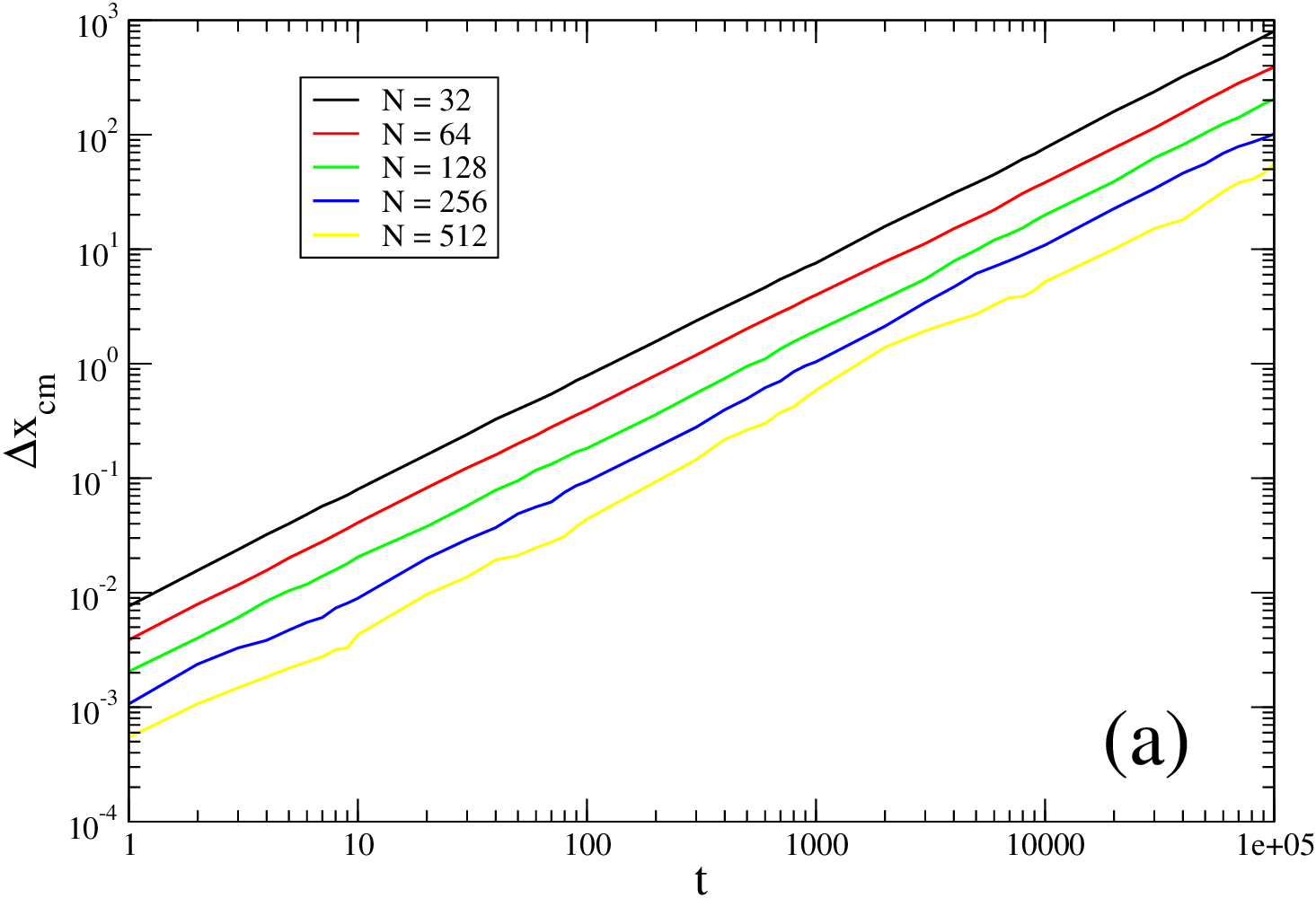}
\includegraphics[clip=true,width=10cm]{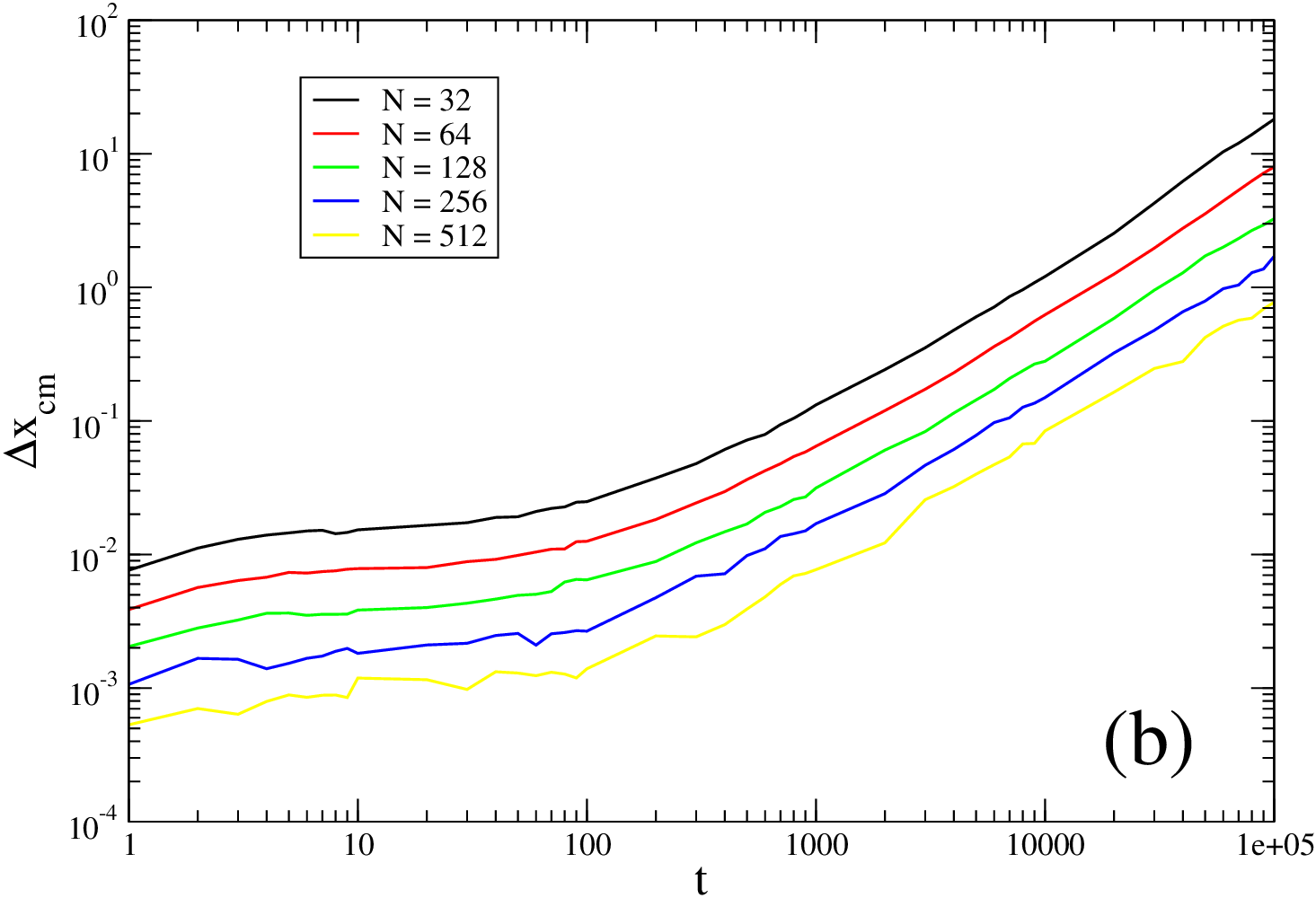}
\end{center}
\caption{Mean-square displacement of the center of mass of the grain boundary,
  $\Delta x_{cm}(t)$ , for Langevin approach simulation without the PN force
  (a) and with the PN force (b). $\Delta x_{cm}(t)$ for $N=32,64,128,256,512$
  is displayed in Log-Log scale.}
\label{graphic4} 
\end{figure}

In what follows, we analyze the dynamical behavior of the stable GB for
$a=b=3\pi$, $A_{PN}=0,0.4$, $A_{PK}=0.14$ and $D=0.25$. In Fig.~\ref{graphic4}
the average position of the GB center of mass $\Delta
x_{cm}(t)$ is displayed with and without the PN force in Log-Log scale for
$N=32,64,128,256,512$.
\begin{figure}[h!]
\begin{center}
\includegraphics[clip=true,width=10cm]{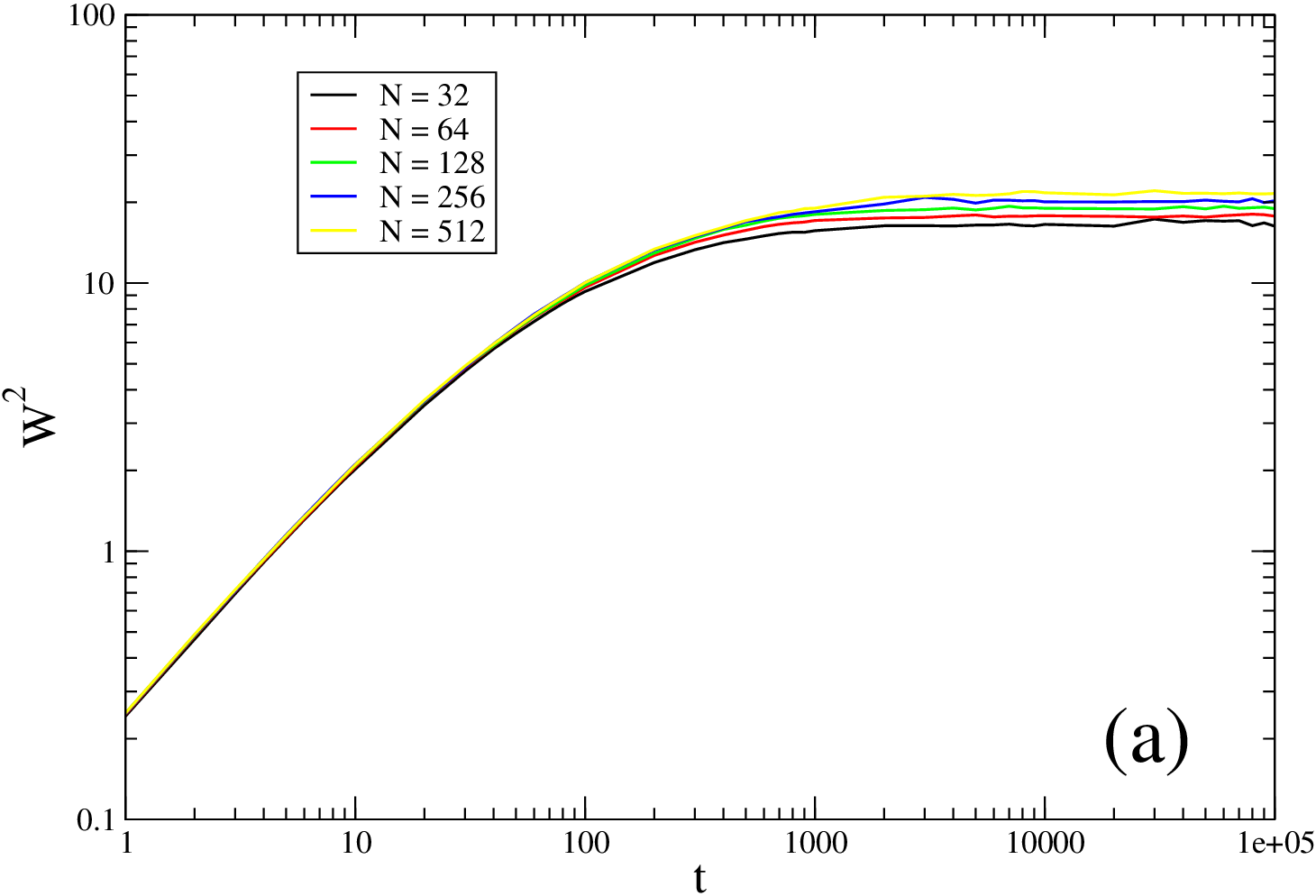}
\includegraphics[clip=true,width=10cm]{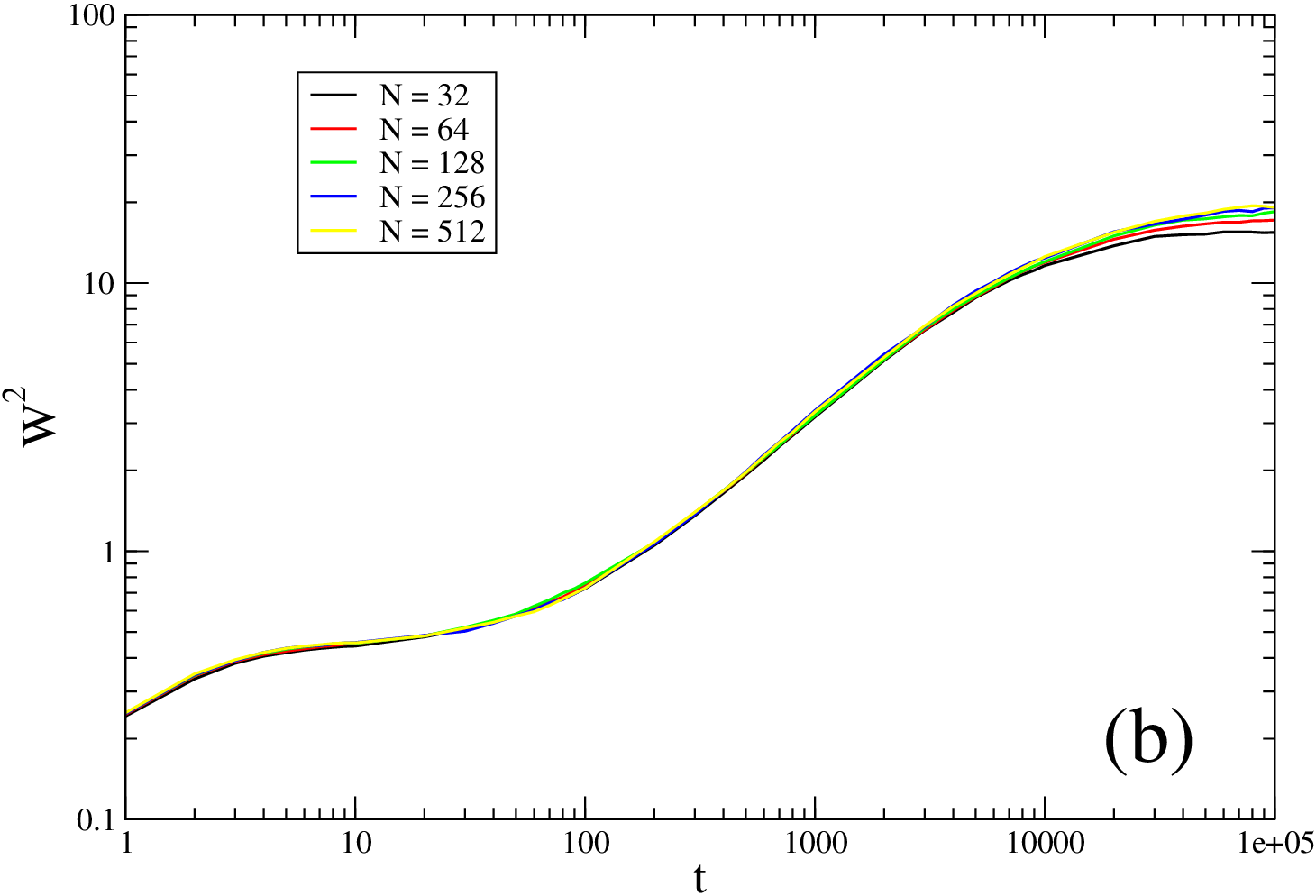}
\end{center}
\caption{Mean-square width of the grain boundary, $W^2(t)$, for Langevin
  approach simulation without the PN force (a) and with the PN force
  (b). $W^2(t)$ for $N=32,64,128,256,512$ is displayed in Log-Log scale.}
\label{graphic2} 
\end{figure}

For long times in both cases we have a linear behavior $\Delta x_{cm}(t)\sim t$, but for short
times, in the presence of the PN force, there is a clear deviation from linearity. 
This result confirms the conclusion
made in the previous section, that the PN force is the cause for
the deviation from linearity of $\Delta x_{cm}(t)$ for short times observed in
Ref.~\cite{trautt}.
Next we characterize the morphology of the GB through the width $W^2(t)$. In
Fig.~\ref{graphic2} $W^2(t)$ for $N=32,64,128,256,512$ is displayed with and 
without the PN force in Log-Log scale.
In the absence of the PN force (Fig.~\ref{graphic2}a) the time dependence of $W^2(t)$ is
qualitatively similar to the same case but with linearized PK force
discussed in the next section, while in the presence of the PN force,
for $A_{PN}=0.4$ (Fig.~\ref{graphic2}b), $W^2(t)$ exhibit a plateau for intermediate
times.

\section{Continuum Theory}

It is possible to develop an analytic expression in the continuum limit
($a\rightarrow 0$, $N\rightarrow\infty$ and $L=Na=const.$) for short or long
times for $W^2(t)$ in absence of the PN force linearizing the PK
force. 
The equation of motion for $F_{PN}=0$ and $F_{PK}$ linearized is
\begin{equation}\label{continuum_GB}
\dot{x}_n(t)=-\frac{\mu
  b^2}{2\pi(1-\nu)\gamma}\sum_{m=1}^{N}\frac{x_n-x_m}{(y_n-y_m)^2}+\eta_n(t) .
\end{equation}

To obtain the short time behavior is sufficient to rewrite
Eq.~\ref{continuum_GB} as a generalized Ornstein-Uhlenbeck process \cite{risken}
\begin{equation}\label{O-U}
\dot{x}_n(t)=\sum_{m=1}^{N}g_{nm}x_m + \eta_n(t) .
\end{equation}
The general solution of Eq.~\ref{O-U} is
\begin{equation}
x_n(t)=\int_{0}^{\infty}\sum_{m=1}^{N}G_{nm}(t')\eta_m(t-t')dt' ,
\end{equation}
with
$\{G_{nm}(t)\}=\hat{G}(t)=e^{\hat{g}t}=\mathbbm{1}+\hat{g}t+\hat{g}^2t^2/2+...$
(where $\mathbbm{1}=\{\delta_{ij}\}$).
From the definition of $W^2(t)$, results
\begin{equation}\label{W2}
W^2(t)=\frac{D}{N}\sum_{n,m=1}^{N}\int_{0}^{t}G_{nm}^2(t')dt'-\frac{D}{N^2}
\sum_{n,m,l=1}^{N}\int_{0}^{t}G_{nm}(t')G_{lm}(t')dt' .
\end{equation}
Replacing the Taylor expansion of the $\hat{G}$ matrix in Eq.~\ref{W2}
one obtains for short times ($t\ll1/\|\hat{g}\|$) that $W^2(t)=(1-1/N)Dt+o(t)$ and
in the continuum limit $W^2(t)=Dt+o(t)$.
To obtain the long times behavior of $W^2(t)$, we rewrite
Eq.~\ref{continuum_GB} in Fourier space \cite{moretti,chui}. Employing the decomposition
$x_m=1/L\sum_k exp(-ikam)x_k$, we obtain
\begin{equation}\label{continuum_fourier}
\fl \dot{x}_k=-\frac{\mu b^2}{2\pi(1-\nu)\gamma a^2L}
\sum_{m=-\infty}^{\infty}e^{ikam} \sum_{n=-\infty}^{\infty}
\frac{\sum_{k'}(e^{-ik'am}-e^{-ik'an})x_{k'}}{(m-n)^2}+\eta_k .
\end{equation} 
The first term in the right-hand side of Eq.~\ref{continuum_fourier} can be
rewritten as
\begin{equation}
-\frac{\mu b^2}{2\pi(1-\nu)\gamma
  a^2L}\sum_{k'}x_{k'}\sum_{m=-\infty}^{\infty}e^{i(k-k')am}
(\sum_{d=-\infty}^{\infty}\frac{1-e^{ik'ad}}{d^2}) ,
\end{equation} 
where $d=m-n$. Using the following results
\begin{equation}
\sum_{d=1}^{\infty}\frac{1}{d^2}=\frac{\pi^2}{6} ,  \ \ \ \ \sum_{d=1}^{\infty}
\frac{cos(cd)}{d^2}=\frac{\pi^2}{6}-\frac{\pi|c|}{2}+\frac{c^2}{4} ,
\end{equation} 
we obtain
\begin{equation}
\sum_{d=-\infty}^{\infty}\frac{1-e^{ik'ad}}{d^2}=2\sum_{d=1}^{\infty}\frac{1-\cos(k'ad)}{d^2}
=\pi|k'|a-\frac{k'^2a^2}{2} ,
\end{equation} 
so that Eq.~\ref{continuum_fourier} becomes
\begin{equation}
\dot{x}_k=-\frac{\mu b^2}{2\pi(1-\nu)\gamma
  a^2}(\pi|k|-\frac{k^2a}{2})x_k+\eta_k .
\end{equation} 
In the long time limit (large $x$, small $k$) the $k^2$ term can be neglected.
Finally in the continuum limit we replace $y_n$, $y_m$ by the continuum variables
$y$, $y'$ and $\langle\eta(y,t)\eta(y',t')\rangle=aD\delta(y-y')\delta(t-t')$.

\begin{figure}[h!]
\begin{center}
\includegraphics[clip=true,width=10cm]{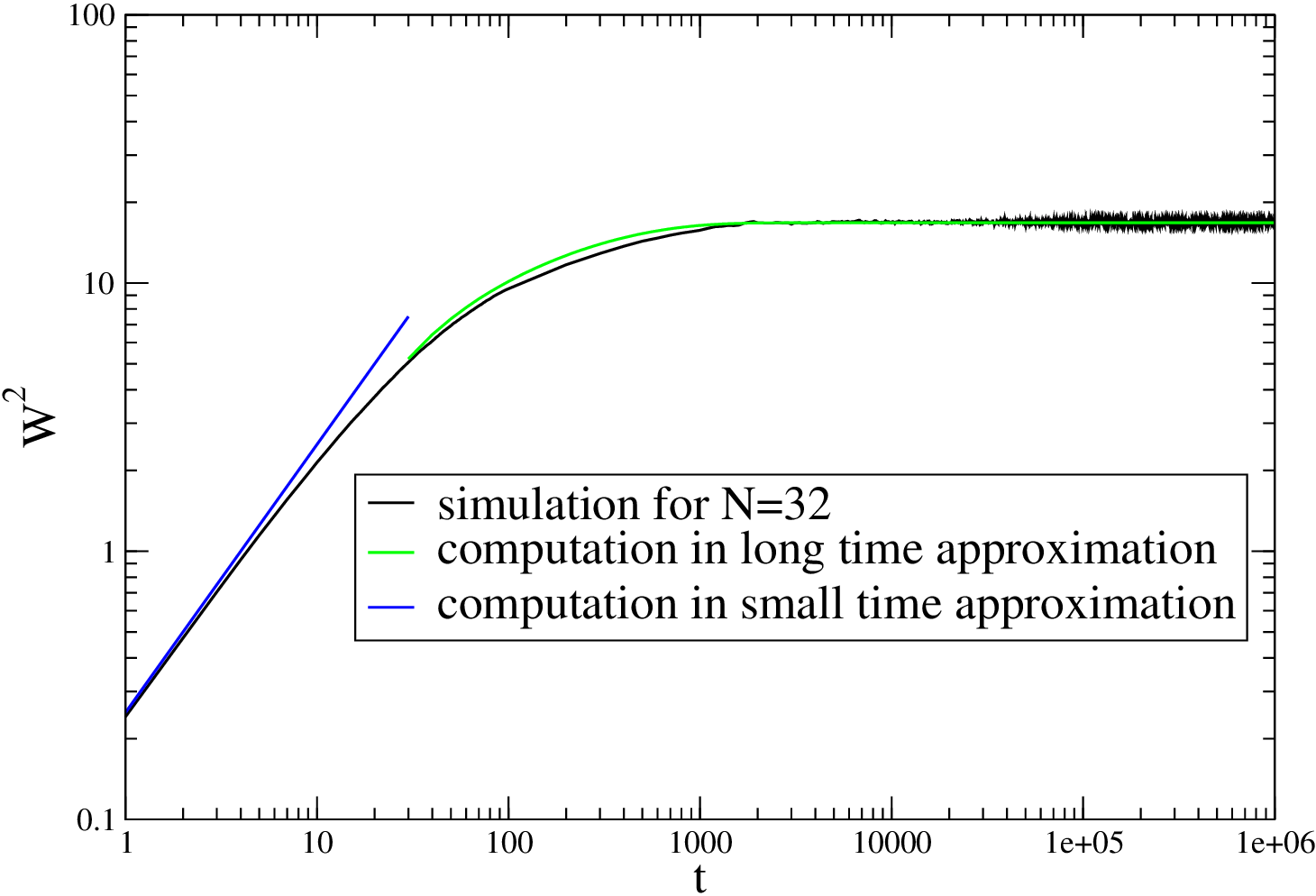}
\end{center}
\caption{Comparison between the mean-square width $W^2(t)$ for linearized
  $F_{PK}$ and $F_{PN}=0$ in the case of $N=32$ computed by
  numerical simulation and the continuum theoretical prediction in the short and
  long times limit. The black line represents the simulated data, the green
  line the long times theoretical prediction with fitted parameters and the blue
  line the short times theoretical prediction.}
\label{graphic6} 
\end{figure}

\begin{figure}[h!]
\begin{center}
\includegraphics[clip=true,width=10cm]{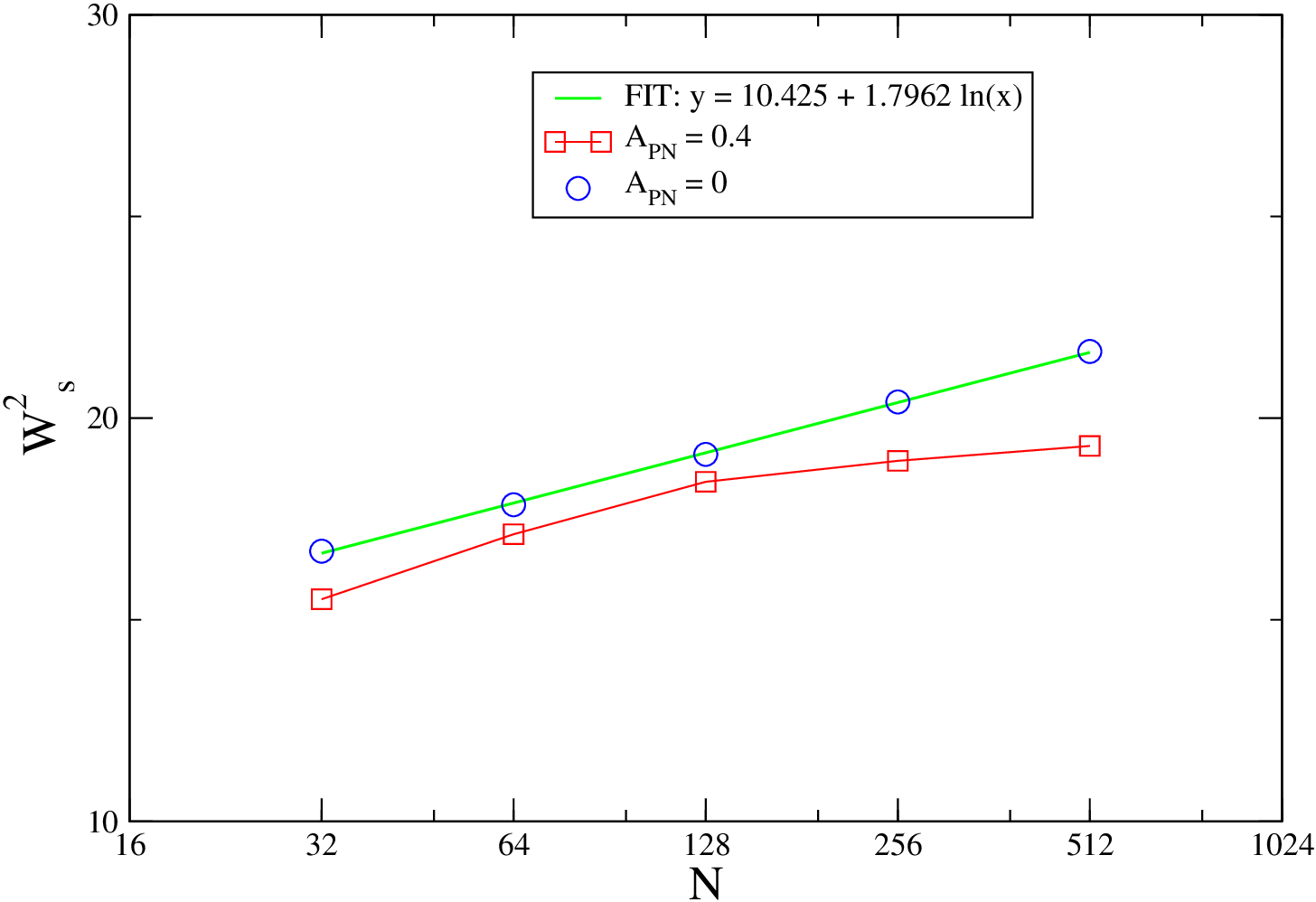}
\end{center}
\caption{Size dependence for the saturation value of the mean-square width
  $W^2_s$ in Log scale for the abscissa (x-axis). As can be seen in figure, in
  the case in which the PN force is absent ($A_{PN}=0$) the relation is
  logarithmic ($W^2_s\sim\log N$), while in presence of the PN force
  ($A_{PN}\ne 0$) a slight deviation from the logarithmic dependence is observed.}
\label{graphic7} 
\end{figure}

Thus Eq.~\ref{continuum_fourier} becomes
\begin{equation}\label{k}
\dot{x}_k=-\alpha|k|x_k+\eta_k ,
\end{equation} 
where $\alpha=\mu b^2/[2(1-\nu)\gamma a^2]$.
Eq.~\ref{k} can be solved exactly and $W^2(t)$ is given by \cite{krug,krug2}
\begin{equation}\label{W_exact}
W^2(t)=\frac{aD}{2\pi\alpha}[\ln(\frac{L}{a})+\ln(1-e^{-4\pi\alpha t/L})] .
\end{equation} 

To reproduce the continuum limit by simulations of Eq.~\ref{continuum_GB} we
would need a very large GB, with $N\gg 512$. Thus a 
comparison between Eq.~\ref{W_exact} and the 
simulations for small $N$ is possible only by introducing some 
effective parameters in Eq.~\ref{W_exact}.
In Fig.~\ref{graphic6} $W^2(t)$ computed by the simulations with $N=32$
(for which the better statistic is available) is compared with the fitted
theoretical prediction for short and long times.  
Eq.~\ref{W_exact} also predicts that the saturated value of width ($W^2_s$) exhibits 
a logarithmic dependence on the GB length $L=N/a$:
$W^2_s\sim\log N$. In the case in which $F_{PK}$ is not linearized this result
is also confirmed by numerical simulations, showing
that $W_s^2$ increases logarithmically with $N$ when $F_{PN}=0$ (see Fig.~\ref{graphic7}). 
In presence of a periodic potential ($F_{PN}> 0$), however, we observe a deviation 
from the logarithmic growth at large $N$. This suggests that the Peierls-Nabarro potential
may set a limit to the GB roughness.

\section{Summary and Discussion}
We have investigated the diffusion of a regularly spaced low-angle grain
boundary in a crystalline material. A typical computational method to describe
the dynamics of the grain boundary is to perform deterministic molecular
dynamics simulations with appropriate interatomic interactions
\cite{trautt}. Here we have employed the over-damped Langevin approach to
obtain a long time description of the dynamics, but in particular to perform a
comparison with molecular dynamics simulations for a specific material \cite{trautt}.
The first results is the interpretation of the early times behavior of the
mean-square displacement $\Delta x_{cm}(t)$. The deviation for early times of
$\Delta x_{cm}(t)$ by the case of the renormalized Brownian motion, that holds
for long times, can be interpreted as the effect on the dislocations of the
periodicity of the lattice giving rise to the Peierls-Nabarro potential.
Secondly the description of the dynamic ($\Delta x_{cm}(t)$) and the morphology
($W^2(t)$) of the grain boundary by means of over-damped Langevin equations is
in qualitatively good agreement with its behavior in real materials, so this
approach can be considered an useful tool for these studies.  
 
\ack
We thank P. Moretti for useful discussions.

\end{document}